\documentclass[a4paper, 10pt, conference]{ieeeconf}      
\usepackage{amsfonts}
\usepackage{amssymb}
\usepackage{amsmath}                                                        
\usepackage{subfig}
\usepackage{epsfig}
\usepackage{graphicx}
\usepackage{psfrag}
\IEEEoverridecommandlockouts                              

\newtheorem{assumption}{Assumption}                                                   
   
\newtheorem{remark}{Remark}  
\title{\LARGE \bf
A Robust Continuous Time Fixed Lag Smoother for Nonlinear Uncertain Systems}

\author{Obaid Ur Rehman and Ian R. Petersen, 
\thanks{O. Rehman and I. R. Petersen are with school of Engineering and Information Technology, University of New South Wales, Canberra, Australia. Emails are \small{ s.obaid.rehman@gmail.com, i.r.petersen@gmail.com}}%
}
\begin{document}
\newtheorem{definition}{Definition}
\newcommand{\eqdef}{\stackrel{\triangle}{=}}

\maketitle
\thispagestyle{empty}
\pagestyle{empty}

\begin{abstract}
This paper presents a robust fixed lag smoother for a class of nonlinear uncertain systems. A unified scheme, which combines a nonlinear robust estimator with a stable fixed lag smoother, is presented to improve the error covariance of the estimation. The robust fixed lag smoother is based on the use of Integral Quadratic Constraints and minimax LQG control. The state estimator uses a copy of the system nonlinearity in the estimator and combines an approximate model of the delayed states to produce a smoothed signal. In order to see the effectiveness of the method, it is applied to a quantum optical phase estimation problem. Results show significant improvement in the error covariance of the estimator using fixed lag smoother in the presence of nonlinear uncertainty.  
\end{abstract}

\section{INTRODUCTION}

The determination of estimate using past and future data is termed a smoothing problem and has been extensively studied in the literature \cite{Frank_estimation,Meditch_smoothing,Kailath_Survey}. It is proved in the early literature that the smoothing process improves the estimation of a Kalman-Bucy filter and thus may be used in variety of applications. Fixed lag smoothing a smoothing technique used where it is desired to calculate state estimate at a fixed time lag $\delta$ behind the current measurement. That is, a smoothed state estimate, $\hat{x}(t|t+\delta)$ is desired at time $t$, given data up to time $t+\delta$. Such smoothers are very useful in the applications where some delay or latency between measurement and the estimation process can be tolerated. The design of a continuous time fixed lag smoother with optimum performance presents a significant challenge due to the presence of delay which means that asymptotic stability cannot be guaranteed. In order to solve instability issue Chirarattananon and Anderson presented an approach in \cite{Anderson_smoother} which can produce an asymptotically stable suboptimal smoother under some conditions.
  
The suboptimal smoothing problem is a two-stage filtering problem where a standard Kalman filter (Kalman-Bucy filter) is first designed and then it is followed by a smoother which produces smoothed estimates of the system \cite{Anderson_smoother}. However, this scheme can be combined to form a unified scheme for both filter and smoother design. Since a unified scheme can be developed for a fixed lag smoother problem, a logical extension of this problem is to replace the Kalman-Bucy filter with a robust filter which gives robust performance in the presence of uncertainty. In this paper, we present a unified scheme which combines a nonlinear robust filter with fixed-lag suboptimal smoother and produce an asymptotically stable filter in the presence of nonlinear uncertainty. The nonlinear robust filter scheme used in this paper is similar to the one presented in \cite{NGC_Petersen}. However, here we use the infinite horizon case. This approach provides a systematic methodology for constructing a robust nonlinear estimator for a class of uncertain nonlinear systems using a copy of the nonlinearity in the nonlinear estimator design. This approach is based on the minimax LQG theory \cite{IP} and uses Integral Quadratic Constraints (IQCs) to characterize the nonlinearities in the system. 
This robust filter is very useful and its effectiveness is revealed in its application to a phase estimation problem for a quantum optical system. In the phase estimation problem, the robust filter is required to handle the measurement nonlinearity in an effective way and simulation results show that the robust filter works better than the standard Kalman filter. The design of a robust filter for the optical system is presented in \cite{Rehman_NGC} and shows significant advantage over a standard Kalman filter. 
 
\begin{figure}[t]
\hfill
\psfrag{y}{$K$}
\begin{center}
\epsfig{file=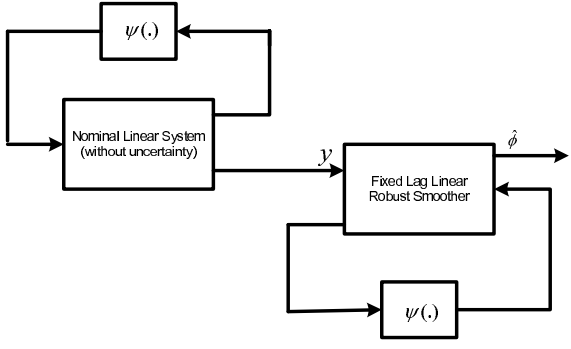, scale=0.6}
\caption{Nonlinear system with nonlinear fixed lag smoother.}
\label{fig:RGCS}
\end{center}
\end{figure}

In this paper, we extend the work of \cite{NGC_Petersen} for the infinite horizon case by including a fixed lag smoother similar to the one used in \cite{Anderson_smoother} and apply the scheme to a quantum optical phase estimation problem (see Fig. \ref{fig:RGCS}). The main contributions of this paper are as follows:
\begin{itemize}
\item Design of an infinite horizon continuous time stable robust fixed lag smoother for a nonlinear uncertain system.
\item Application of the scheme to an adaptive phase estimation problem for a quantum optical system.
\end{itemize}

The paper is organized as follows. Section \ref{sec:system} presents a description of the class of uncertain systems considered in this paper. The development of a fixed lag smoother for the system is presented in Section \ref{sec:SFLS}. The relevant phase estimation problem is solved in Section \ref{sec:PE} and the paper is concluded in Section \ref{sec:concl}.
\section{System Definition}\label{sec:system}
Let us consider a system on certain probability space driven by the noise input $W(\cdot)$ as follows (see also \cite{NGC_Petersen}):

\begin{equation}
\begin{split}
dx(t)&=A x(t) dt+ [\sum_{i=1}^g \bar{B}_{1i}\mu_i (t)+ \sum_{s=1}^k B_{1s} \xi_s(t)] dt\\
&+B_1  dW(t);\quad x(0)=x_0,\\
w(t)&=C_0 x(t);\\
\zeta_1(t)&=C_{1,1} x(t);\\
& \quad \vdots\\
\zeta_k(t)&=C_{1,k} x(t);\\
\nu_1 (t)&=\bar{C}_{1,1} x(t);\\
& \quad \vdots\\
\nu_g (t)&=\bar{C}_{1,g}  x(t);\\
dy(t)&=C_2 x(t) dt+ [\sum_{i=1}^g \bar{D}_{21,i}\mu_i (t)\\
&+ \sum_{s=1}^k D_{21,s} \xi_s(t)] dt+D_{21}  dW(t);\quad y(0)=0,\\
\end{split}
\label{eqsystem}
\end{equation}
where $x(t)\in \Re^{\bar{n}}$ is the state, $w(t)\in \Re^m$ is the estimated output, $\zeta_1(t)\in \Re^{h_1}, \cdots, \zeta_k(t)\in \Re^{h_k}$ are the uncertainty outputs, $\nu_1(t)\in \Re^{h_1}, \cdots, \nu_g(t)\in \Re$ are the nonlinearity outputs, $\xi_1(t)\in \Re^{r_1}, \cdots, \xi_k(t)\in \Re^{r_k}$ are the uncertainty inputs, $\nu_1(t)\in \Re, \cdots, \nu_g(t)\in \Re$ are nonlinearity inputs, and $y(t)\in \Re^l$ is the measured output.

The nonlinearity inputs are related to the nonlinearity outputs by the following nonlinear relations
\begin{equation}
\label{eqnon1}
\mu_i(t)=\psi_i (\nu_i(t)) \quad \forall i=1,2, \cdots, g,
\end{equation}
where the nonlinear functions $\psi_i(\cdot)$ are such that $\psi_i(0)=0$ and satisfy the following global Lipschitz conditions:
\begin{equation}
\label{eqcond1}
\vert \psi_i (\nu_i(t)) -  \psi_i (\nu_i(t)) \vert \mu_i(t) \leq \beta_i \vert \nu_1 - \nu_2 \vert
\end{equation}
for all $\nu_1$, $\nu_2$ and for all $i=1,2, \cdots, g$. The uncertainty in the system is described by the following Stochastic Integral Quadratic Constraint. 
\begin{equation}
\label{eqIQC1}
E\int_0^\infty \Vert \xi_s (t) \Vert ^2 dt \leq E [\int_0^\infty \Vert \zeta_s (t) \Vert ^2 dt +x(0)^T S_s x(0) ].
\end{equation}
\begin{definition}
An uncertainty is an admissible uncertainty for the system (\ref{eqsystem}), (\ref{eqnon1}) if given any strong solution to the system  (\ref{eqsystem}), (\ref{eqnon1}) then 
for all $s=1, \cdots, k$ and $S_s$ are given positive-definite matrices and $\Vert \cdot \Vert$ denotes the standard Euclidean norm.
\end{definition}

\section{Suboptimal Fixed Lag Smoother Design}\label{sec:SFLS}
In this section, we present the development of a fixed lag smoother for the system defined in (\ref{eqsystem}). Let us define the filtered state estimate $\hat{x}_f(t|t)$ of $x(t)$ and the fixed-lag smoothed estimate $\hat{x}_s$ as follows:
\begin{equation}
\begin{split}
\hat{x}_f(t|t)&=E[x(t)|y(\tau), \quad t_0\geq \tau <t]\\
\hat{x}_s(t|t+\delta)&=E[x(t)|y(\tau), \quad t_0\geq \tau <t+\delta]
\end{split}
\end{equation}
In order to delay the estimated signal an approximate model for a delay $\delta$ is required to be augmented with the system (\ref{eqsystem}). The approximation of the delay by a linear finite dimensional system is given as follows:
\begin{equation}
\label{eqdelay}
\begin{split}
\dot{x}_a(t)=F_a x_a+G_a w(t) \\
w_a(t)=H_a x_a+ J_a w(t)
\end{split}
\end{equation}
where $x_a(t)\in \Re^{n_a}$. Also, $F_a$, $Ga$, $H_a$ and $J_a$ are chosen such that (\ref{eqdelay}) approximates the actual delay operator and the approximate system remains asymptotically stable (see \cite{Anderson_smoother}). 
 
Let us combine (\ref{eqsystem}) and (\ref{eqdelay}) to form an augmented system which represents a new system where signal $w(t)$ is delayed by $\delta$.
\begin{equation}
\begin{split}
dx_p(t)&=A_p x_p(t) dt+ [\sum_{i=1}^g \bar{B}_{p1i}\mu_i (t)+ \sum_{s=1}^k B_{p1s} \xi_s(t)] dt\\
&+B_{p1}  dW(t);\\
w(t)&=C_{p0} x_p(t);\\
\zeta_{1}(t)&=C_{p1,1} x_p(t);\\
& \quad \vdots\\
\zeta_{k(t)}&=C_{p1,k} x_p(t);\\
\nu_{1} (t)&=\bar{C}_{p1,1} x_p(t);\\
& \quad \vdots\\
\nu_g (t)&=\bar{C}_{p1,g}  x_p(t);\\
dy(t)&=C_{p2} x_p(t) dt+ [\sum_{i=1}^g \bar{D}_{21,i}\mu_i (t)\\
&+ \sum_{s=1}^k D_{21,s} \xi_s(t)] dt+D_{21}  dW(t);\quad y(0)=0,\\
\end{split}
\label{eqNsystem}
\end{equation}
where $x_p(t)\in \Re^n$ $(n=\bar{n}+n_a)$ and
\[
A_p=\left[
\begin{array}{cc}
A & 0\\
G_a C_o & F_a
\end{array}\right],~
x_p(t)=\left[
\begin{array}{c}
x(t) \\
x_a(t)
\end{array}\right],
\]
\[
\bar{B}_{p1i}=\left[
\begin{array}{c}
\bar{B}_{p1i} \\
0
\end{array}\right],~ \bar{B}_{p1s}=\left[
\begin{array}{c}
\bar{B}_{p1s} \\
0
\end{array}\right] 
\]
\[
{B}_{p1}=\left[
\begin{array}{c}
{B}_{p1} \\
0
\end{array}\right],~C_{p2}=\left[
\begin{array}{cc}
C_{2} & 0
\end{array}\right],~C_{po}=\left[
\begin{array}{cc}
C_{o} & 0
\end{array}\right]
\]
\[
C_{p1,k}=\left[\begin{array}{cc}
C_{1,k} & 0
\end{array}\right]~
\bar{C}_{p1,g}=\left[\begin{array}{cc}
\bar{C}_{1,g} & 0
\end{array}\right]
\]
Note that measurement equation remains same. The delayed signal $w_a(t)$ which is need to be estimated is written as follows:
\begin{equation}
w_a(t)=C_a x_d
\end{equation}
where $C_a=[J_a C_o\quad Ha]$.
The model (\ref{eqdelay}) delays the signal $w(t)$ by amount of $\delta$ as follows:
\begin{equation}
w_a(t+\delta) \approx w(t),
\end{equation}
The optimum filtered estimate of $w_a(t+\delta)$ is an approximation of the fixed-lag smoothed estimate of $w(t)$; i.e.
\begin{equation}
\hat{w}_a(t+\delta)=\hat{w}(t|t+\delta)
\end{equation}
Note that the augmentation of the delay  in (\ref{eqNsystem}) does not affect the signal $w(t)$ and the measurement $y$. This augmentation is only an artifact to solve the smoothing problem under consideration. 
The nonlinear dynamic state estimator with smoother chosen for the system (\ref{eqNsystem}) is as follows:
\begin{equation}
\label{eqestimator}
\begin{split}
d\hat{x}(t)&= (A_c \hat{x}(t)+ \sum_{i=1}^g \bar{G}_{ci} \tilde{\mu}_i(t))dt\\
&+B_c dy; \quad \hat{x}(0)=\hat{x}_0,\\
\tilde{\nu}_1(t)&=\bar{K}_{c1}\hat{x}(t);\\
& \quad \vdots\\
\tilde{\nu}_g(t)&=\bar{K}_{cg}\hat{x}(t);\\
\hat{w}(t)&=C_c  \hat{x}(t),
\end{split}
\end{equation}
where 
\begin{equation}
\label{eqnon2}
\tilde{\mu}_i(t)=\psi_i (\tilde{\nu}_i(t)) ~\text{for} ~i=1,2, \cdots, g. 
\end{equation}

The nonlinear dynamic estimator (\ref{eqestimator}) is designed such that it provides an upper bound on the following cost functional:
\begin{equation}
\label{eqfcost}
J(\hat{w}(\cdot))=E\big[\frac{1}{2}\int_0^\infty \Vert \hat{w}_a(t|t+\delta)-w_a(t) \Vert^2dt\big].
\end{equation}
The problem of smoothed state estimation is to use the system (\ref{eqestimator}) to find a state estimate for the system (\ref{eqNsystem}) by considering the cost functional (\ref{eqfcost}). This problem can be transformed to a linear control problem for the system (\ref{eqNsystem}) with repeated nonlinearities by using the following compact form of the estimator (\ref{eqestimator}).
\begin{equation}
\label{eqcontrol}
\begin{split}
d\hat{x}_p(t)&=A_c \hat{x}_p(t) dt+ \tilde{B_c} d\tilde{y}(t),\\
\tilde{u}(t)&=\tilde{C}_c \hat{x}_p(t),
\end{split}
\end{equation}
where 
\[
\tilde{y}(t)\eqdef
\left[
\begin{array}{c}
y(t) \\
\tilde{\mu}_1(t)\\
\vdots\\
\tilde{\mu}_g(t)
\end{array}\right];\quad
\tilde{u}(t)\eqdef
\left[
\begin{array}{c}
\hat{w}(t) \\
\tilde{\nu}_1(t)\\
\vdots\\
\tilde{\nu}_g(t)
\end{array}\right];
\]
\[
\tilde{B}_c \eqdef \left[
\begin{array}{cccc}
B_c & \bar{G}_{c1} & \cdots & \bar{G}_{cg}
\end{array}\right];~\tilde{C}_c \eqdef
\left[
\begin{array}{c}
C_c \\
\bar{K}_{c1}\\
\vdots\\
\bar{K}_{cg}
\end{array}\right].
\]
The main idea here is to characterize the repeated nonlinearities (\ref{eqnon1}) and (\ref{eqnon2}) by certain IQCs. The condition (\ref{eqcond1}) for both (\ref{eqnon1}) and (\ref{eqnon2}) imply that the following IQCs are satisfied (see \cite{NGC_Petersen} for details):
\begin{equation}
\label{eqIQC2}
\begin{split}
E &\int_0^\infty [\mu_i(t)-\tilde{\mu}_i(t)]^2 dt  \\ &\leq E [\int_0^\infty \beta_i^2[\nu_i-\tilde{\nu}(t)]^2 dt+x(0)^T S_{1i} x(0)],
\end{split}
\end{equation}
\begin{equation}
\label{eqIQC3}
\begin{split}
E &\int_0^\infty [\mu_i(t)]^2 dt \\ &\leq E [\int_0^\infty \beta_i^2[\nu_i]^2 dt+x(0)^T S_{2i} x(0)],
\end{split}
\end{equation}
\begin{equation}
\label{eqIQC4}
\begin{split}
E &\int_0^\infty [\tilde{\mu}_i(t)]^2 dt \\ &\leq E [\int_0^\infty \beta_i^2[\tilde{\nu}(t)]^2 dt+x(0)^T S_{3i} x(0)]
\end{split}
\end{equation}
for all $i=1, \cdots , g$. Here the $\bar{S}_{1i}$, $\bar{S}_{2i}$, $\bar{S}_{3i}$ are any positive definite matrices.

The system (\ref{eqNsystem}) can be written in a compact form as follows:
\begin{equation}
\label{eqCsystem}
\begin{split}
dx_p(t)&=[A_p x_p(t)+\tilde{B}_1  \tilde{\xi}(t)] dt + B_{p1}  dW(t);\\
w(t)&=C_{p0}  x_p(t);\\
\tilde{\zeta}(t)&=\tilde{C}_1  x_p(t)+\tilde{D}_{12} \tilde{u}(t);\\
d\tilde{y}(t)&=[\tilde{C}_2 x_p(t)+\tilde{D}_{21} \tilde{\xi}(t)]dt +\bar{D}_{21} dW(t),
\end{split}
\end{equation}
where
\[
\tilde{\xi}(t)=
\left[
\begin{array}{c}
\tilde{\xi}_1 (t)\\
\vdots\\
\tilde{\xi}_{k+2g}(t)
\end{array}\right]\eqdef \left[
\begin{array}{c}
\xi_1 (t)\\
\vdots\\
\xi_k (t)\\
\mu_1 (t)\\
\vdots\\
\mu_g (t)\\
\tilde{\mu}_1 (t)\\
\vdots\\
\tilde{\mu}_g (t)
\end{array}\right];
\]
\[
\tilde{\zeta}(t)=
\left[
\begin{array}{c}
\tilde{\zeta}_1 (t)\\
\vdots\\
\tilde{\zeta}_{k+2g}(t)
\end{array}\right]\eqdef \left[
\begin{array}{c}
\zeta_1 (t)\\
\vdots\\
\zeta_k (t)\\
\mu_1 (t)\\
\vdots\\
\mu_g (t)\\
\tilde{\nu}_1 (t)\\
\vdots\\
\tilde{\nu}_g (t)
\end{array}\right];
\]
\[
\tilde{B}_1 = \left[
\begin{array}{ccccccc}
B_{p1,1} & \cdots & B_{p1,k} &  \bar{B}_{p1,1} & \cdots   \bar{B}_{p1,g} & 0_{n \times g}
\end{array}\right];
\]

\[
\tilde{C}_1=\left[
\begin{array}{c}
C_{p1,1}\\
\vdots\\
C_{p1,k}\\
\bar{C}_{p1,1}\\
\vdots\\
\bar{C}_{p1,g}\\
0_{g \times n}\\
\end{array}\right];~ \tilde{D}_{12}=\left[
\begin{array}{cc}
0_{h_1 \times m} & 0_{h_1 \times g}\\
\vdots & \vdots \\
0_{h_k \times m} & 0_{h_k \times g}\\
0_{1 \times m} & 0_{1 \times g}\\
\vdots & \vdots \\
0_{1 \times m} & 0_{1 \times g}\\
0_{g \times m} & I_{g \times g}\\
\end{array}\right];
\]
\[
\tilde{C}_{2}=\left[
\begin{array}{c}
C_{p2}\\
0_{g \times n}\\
\end{array}\right]
;~ \bar{D}_{21}=\left[
\begin{array}{c}
D_{21}\\
0_{g \times (h+g)}~J_{21}\\
\end{array}\right];
\]
\[
\tilde{D}_{21}=\left[
\begin{array}{ccccccc}
D_{21,1} & \cdots& D_{21,k} & \bar{D}_{21,1} & \cdots& \bar{D}_{21,g}& 0_{l \times g}\\
0_{g \times r_1} & \cdots& 0_{g \times r_k} & 0_{g \times 1} & \cdots& 0_{g \times 1}& I_{g \times g}\\
\end{array}\right].
\]
Also $h=\sum_{i=1}^k h_i$, $r=\sum_{i=1}^k r_i$, $p=h+2g$ and $J_{21}$ is any suitable matrix satisfying $J_{21}>0$.
Considering these new variables, the IQCs (\ref{eqIQC1}), (\ref{eqIQC2}), (\ref{eqIQC3}), and (\ref{eqIQC4}) can be written as follows:
\begin{small}
\begin{equation}
\label{eqIQC}
E \int_0^\infty \tilde{\xi}^T(t) \tilde{M}(\lambda) \tilde{\xi}(t) dt \leq E[\int_0^\infty \tilde{\zeta}^T(t) \tilde{N} \tilde{\zeta}(t)+x(0)^T \tilde{S}_i x(0)]
\end{equation}
\end{small}
for all $\lambda=[\lambda_1~\lambda_2,\cdots, \lambda_{\tilde{k}}]^T\in \Re^{\tilde{k}}$ where $\tilde{k}=k+3g$. Also,  $\tilde{M}(\lambda)=\sum_{i=1}^{\tilde{k}}=\lambda_i M_i\geq 0$ and $\tilde{N}(\lambda)=\sum_{i=1}^{\tilde{k}}=\lambda_i N_i\geq 0$ where $M_i=m_i^T m_i\geq 0$, $N_i=n_i^T n_i\geq 0$ and $\tilde{S}_i$ are positive-definite matrices.

The constraint on the $\tilde{M}(\lambda)$ is defined as follows:
\begin{equation}
\label{eqcons}
\lambda\in \tilde{\Gamma}: \tilde{M}(\lambda)^{-1} \geq J J^T~ \forall t,
\end{equation}
where $\tilde{\Gamma}=\{\lambda\in \Re^{\tilde{k}}:\lambda_i\geq0~ \forall i \tilde{M}(\lambda)>0\}$.
\begin{assumption}
There exists a square matrix function $J$ such that we can write
\begin{equation}
\left[
\begin{array}{c}
B_{p1} \\
\bar{D}_{21} \\
\end{array}\right]=\left[
\begin{array}{c}
\tilde{B}_{1} \\
\tilde{D}_{21} \\
\end{array}\right]J.
\end{equation}
\end{assumption}

\begin{assumption}
There exist a constant $d_0>0$ such that 
\begin{equation}
\bar{D}_{21}\bar{D}_{21}^T=\tilde{D}_{21} JJ^T \tilde{D}_{21}^T\geq d_0 I
\end{equation}
for all $t$.
\end{assumption}

\begin{assumption}
There exist a constant $\bar{\tau}>0$ such that the following conditions hold:
\begin{enumerate}
\item  The algebraic Riccati equation
\begin{equation}
\begin{split}
&(A_p-\tilde{B}_1 \tilde{M}(\lambda)^{-1}\tilde{D}_{21}^T E_\lambda^{-1}\tilde{C}_2)^T Y\\
&+Y(A_p-\tilde{B}_1 \tilde{M}(\lambda)^{-1}\tilde{D}_{21}^T E_\lambda^{-1}\tilde{C}_2)\\
&-Y(\tilde{C}_2^T E_\lambda^{-1} \tilde{C}_2-\frac{1}{\bar{\tau}}R_{\bar{\tau},\lambda})Y+\tilde{B}_1 \tilde{M}(\lambda)^{-1} \tilde{B}_1^T\\
&- \tilde{B}_1 \tilde{M}(\lambda)^{-1} \tilde{D}_{21}^T E_\lambda^{-1} \tilde{D}_{21} \tilde{M}(\lambda)^{-1} \tilde{B}_1^T=0
\end{split}
\end{equation}
has a symmetric positive definite solution.
\item The algebraic Riccati equation
\begin{equation}
\begin{split}
&XA_p+A_p^TX+(R_{\bar{\tau},\lambda}-\Gamma_{\bar{\tau},\lambda})^{-1} \Gamma_{\bar{\tau},\lambda}^T\\
&+\frac{1}{\bar{\tau}} X \tilde{B}_1 \tilde{M}(\lambda)^{-1} \tilde{B}_1^T X=0
\end{split}
\end{equation}
has a symmetric nonnegative definite solution.
\item and
\begin{equation}
\rho(Y X)<\bar{\tau}
\end{equation}
\end{enumerate}
where $\rho(\cdot)$ denotes the spectral radius of a matrix and 
\[
R_{\bar{\tau},\lambda}=C_{p0}^T C_{p0}+\bar{\tau} \tilde{C}_1^T\tilde{N}(\lambda)\tilde{C}_1,
\]
\[
G_{\bar{\tau},\lambda}=\left[
\begin{array}{cc}
I_{m \times m} & 0_{m \times g}\\
0_{g \times m} & 0_{g \times g}
\end{array}\right]+\bar{\tau} \tilde{D}_{12}^T \tilde{N}(\lambda) \tilde{D}_{12},
\]
\[
\Gamma_{\bar{\tau},\lambda} \eqdef -[C_{p0}^T\quad 0_{n \times g}]+ \bar{\tau} \tilde{C}_{1}^T \tilde{N}(\lambda) \tilde{D}_{12}.
\]
Also, $\tilde{M}(\lambda)=\sum_{i=1}^{\bar{k}} \lambda_i M_i \geq 0$ where $\lambda=[\lambda_1~\lambda_2~\cdots \lambda_{\tilde{k}}]^T\in \Re^{\tilde{k}}$.
\end{assumption}

If all the given assumptions are satisfied then filter and smoother gain are calculated as follows:
\begin{equation}
\tilde{B}_c=(Y \tilde{C}_2^T+\tilde{B}_1 \tilde{M}(\lambda)^{-1} \tilde{D}_{21}^T)E_{\lambda}^{-1};
\end{equation}
The gain $\tilde{B}_c$ consists of two components of gains. The first component is the nonlinear robust filter gain of size $\bar{n} \times g$ and other component is the smoother gain of size $n_a \times g$. The remaining parameters in (\ref{eqcontrol}) can be determined using the following relations (see \cite{IP}):
\begin{equation}
\label{eqCparams}
\begin{split}
A_c&=A+\frac{1}{\bar{\tau}}Y R_{\bar{\tau},\lambda}-(Y \tilde{C}_2^T+\tilde{B}_1 \tilde{M}(\lambda)^{-1} \tilde{D}_{21}^T)E_{\lambda}^{-1} \tilde{C}_2\\
&- \frac{1}{\bar{\tau}} Y \Gamma_{\bar{\tau},\lambda} G_{\bar{\tau},\lambda}^{-1} \Gamma_{\bar{\tau},\lambda}^T(I-\frac{1}{\bar{\tau}} Y X)^{-1};\\
\tilde{C}_c&=-G_{\bar{\tau},\lambda}^{-1} \Gamma_{\bar{\tau},\lambda}^T [I-\frac{1}{\bar{\tau}} Y X]^{-1}.
\end{split}
\end{equation}

The corresponding guaranteed cost bound is given by the following relation.
\begin{equation}
\label{eqbound}
\begin{split}
V_{\bar{\tau}}=\frac{1}{2}& tr[Y R_{\bar{\tau},\lambda}+(Y \tilde{C}_2^T+\tilde{B}_1 \tilde{M}(\lambda)^{-1} \tilde{D}_{21}^T)E_\lambda^{-1} \\
&\times (\tilde{C}_2 Y + \tilde{D}_{21} \tilde{M}(\lambda)^{-1} \tilde{B}_1^T) \times X(I-\frac{1}{\bar{\tau}}Y X)^{-1}].
\end{split}
\end{equation}

\subsection{Error covariance}
Let us define the error covariance of the suboptimal smoothed estimate as follows:
\begin{equation}
\label{eqErrorco}
P_{sa}(t|t+\delta)=E \left\{[\hat{w}_a(t+\delta)-w(t)][\hat{w}_a(t+\delta)-w(t)]\right\}
\end{equation}
In order to find $P_{sa}(t|t+\delta)$ it is required to write closed loop system using (\ref{eqNsystem}) and (\ref{eqcontrol}). The uncertainty input can be written as $\tilde{\xi}=\Delta\tilde{\zeta}$ where $\vert\Delta\vert\leq 1$.

\begin{equation}
\begin{split}
dx_p(t)=[A_p x_p(t)+\tilde{B}_1 [\Delta (\tilde{C}_1  x_p(t)&+\tilde{D}_{12} \tilde{u}(t))]] dt\\
 + B_{p1}  dW(t);
\end{split}
\end{equation}
\begin{equation}
\begin{split}
dx_p(t)=[(A_p+\tilde{B}_1 \Delta \tilde{C}_1) x_p(t) +\tilde{B}_1 \Delta \tilde{D}_{12} \tilde{C}_c \hat{x}_p  ] dt &\\
+ B_{p1}  dW(t);
\end{split}
\end{equation}
Also, (\ref{eqcontrol}) can be written as follows:
\begin{small}
\begin{equation}
\begin{split}
d\hat{x}_p(t)&=[A_c \hat{x}_p(t)+\tilde{B}_c(\tilde{C}_2x_p(t)+\tilde{D}_{21}\tilde{\xi}(t))]dt+\tilde{B}_c\bar{D}_{21}(t)]dW(t)\\
d\hat{x}_p(t)&=[A_c \hat{x}_p(t)+\tilde{B}_c\tilde{C}_2 x_p (t)+\tilde{B}_c\tilde{D}_{21}\Delta(\tilde{C}_1x_p(t)\\
&+\tilde{D}_{12}\tilde{C}_c \hat{x}_p(t))]dt
+\tilde{B}_c\bar{D}_{21}(t)dW(t),
\end{split}
\end{equation}
\end{small}
and finally,
\begin{equation}
\begin{split}
d\hat{x}_p&=[(A_c+\tilde{B}_c\tilde{D}_{21}\Delta\tilde{D}_{12}\tilde{C}_c)\hat{x}_p\\
&+(\tilde{B}_c\tilde{C}_2+\tilde{B}_c\tilde{D}_{21}\Delta \tilde{C}_1)x_p(t)]dt+\tilde{B}_c\bar{D}_{21}(t)dW(t)
\end{split}
\end{equation}
The augmented closed loop system is given below.
\begin{equation}
\label{eqclosed}
dX=\mathbf{A} X dt+ \mathbf{B} dW,
\end{equation}
where
\[
dX=\left[
\begin{array}{c}
dx_p(t) \\
 d\hat{x}_p(t)
\end{array}
\right],\quad X=\left[
\begin{array}{c}
x_p(t) \\
 \hat{x}_p(t)
\end{array}
\right],
\]
\[
\mathbf{A}=\left[
\begin{array}{cc}
 A_p+\tilde{B}_1 \Delta \tilde{C}_1 & \tilde{B}_1 \Delta \tilde{D}_{12}\tilde{C}_c  \\
 \tilde{B}_c\tilde{C}_2+\tilde{B}_c\tilde{D}_{21}\Delta \tilde{C}_1 & A_c+\tilde{B}_c\tilde{D}_{21}\Delta\tilde{D}_{12}\tilde{C}_c \\
 \end{array}
\right],
\]
\[
\mathbf{B}=\left[
\begin{array}{c}
[B_{p1}~0] \\
\tilde{B}_c\bar{D}_{21}
\end{array}
\right].
\]
The error covariance (\ref{eqErrorco}) can be expanded to yield following relation:
\begin{equation}
\label{eqCo}
\begin{split}
&P_{sa}(t|t+\delta)=[C_{po}\quad 0]\mathbf{P}(t,t)[C_{po}\quad 0]'\\
&-[0\quad C_a]\mathbf{P}(t+\delta,t)[C_{po}\quad 0]'-[C_{po}\quad 0]\mathbf{P}(t+\delta,t)[0\quad C_a]'\\
&+[0\quad C_a]\mathbf{P}(t+\delta,t+\delta)[0\quad C_a]',
\end{split}
\end{equation}

where $\mathbf{P}(t,\tau)=E[X(t)X(\tau)]$ is the steady state error covariance matrix which can be obtained by using following Lyapunov equation for the stationary system:
\begin{equation}
\mathbf{A} \mathbf{P}(t,t)+\mathbf{P}(t,t) \mathbf{A}'+\mathbf{B}\mathbf{B}'=0.
\end{equation}
Also, $\mathbf{P}(t+\delta,t)=\Phi(\delta) \mathbf{P}(t,t)$ and $\mathbf{P}(t+\delta,t+\delta)=\mathbf{P}(t,t)$, where $\Phi(\delta)$ is transition matrix associated with (\ref{eqclosed}):
\begin{equation}
\Phi(\delta)=e^{\mathbf{A} \delta}.
\end{equation}
\begin{figure}[t]
\hfill
\begin{center}
\epsfig{file=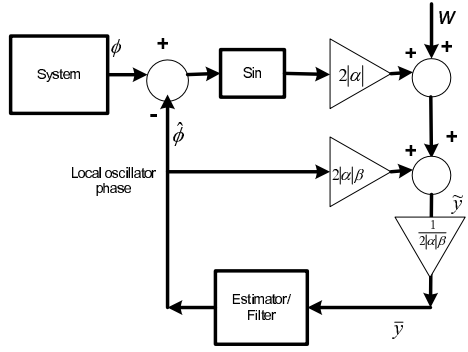, scale=0.6}
\caption{Adaptive homodyne system with a robust nonlinear filter.}
\label{fig:closed_homodyne}
\end{center}
\end{figure}
\section{Example}\label{sec:PE}
In this section, the scheme developed above for the fixed lag smoother is applied to a phase estimation problem for a quantum optical system. This example is taken from \cite{Wheatley_Adaptive} where an adaptive homodyne technique is applied to estimate the phase by linearizing the system. Later in another work \cite{Shibdas_RF}, it is shown that the error covariance of this phase estimation error can be improved by using a linear robust filter in the feedback loop  (see Fig. \ref{fig:closed_homodyne}). In our previous work \cite{Rehman_RNE}, we suggested a nonlinear robust filter, which shows significant improvement over the linear filter. This is due to the fact that a nonlinear uncertainty is present in the measurement and a nonlinear robust filter effectively deals with the nonlinearity in the system. Here, in this example, we show that by using the unified scheme for nonlinear fixed lag smoothing presented in Section \ref{sec:SFLS}, the error covariance improves significantly. The governing equation for this system can be written in the form of a stochastic differential equation as follows:
\begin{equation}
\label{eqdphi}
d\phi(t)=-\lambda \phi(t) dt + \sqrt{\kappa} dV(t).
\end{equation}
where $dV(t)$ is a Wiener increment satisfying $(dV)^2=dt$. The photon current for a continuous coherent input beam is given by
\begin{equation}
I(t)dt=2 Re(\alpha e^{-i \Phi(t)} )dt + dW(t),
\end{equation}
where $\Phi(t)= \hat{\phi}+\pi/2$ is the phase of the local oscillator, and $dW(t)$ is a Wiener increment independent of $dV(t)$. The homodyne photocurrent is given by the following relation:
\begin{equation}
\label{eqIt}
I(t)dt=2 \vert \alpha \vert \sin[\phi - \hat{\phi}(t)] dt + dW(t).
\end{equation}
\begin{figure}[t]
\hfill
\begin{center}
\epsfig{file=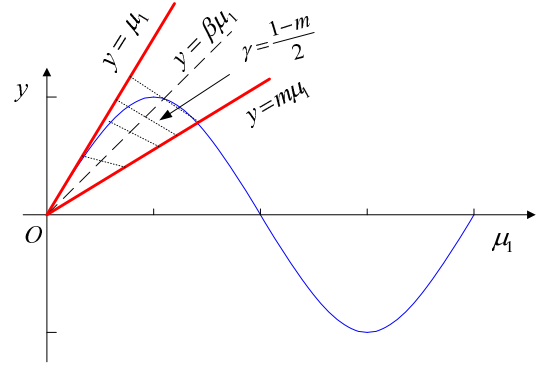, scale=0.5}
\caption{Sector bound.}
\label{fig:sector_bound}
\end{center}
\end{figure}
\subsection{Uncertainty Modelling}\label{sec:UM} 
The measurement equation (\ref{eqIt}) of the optical system is a nonlinear equation. However, we can approximate the model (\ref{eqdphi}), (\ref{eqIt})   with a  linear system having a sector bounded nonlinear uncertainty. Let us define $\mu_1:=\phi-u$ where $u=\hat{\phi}$ is the controller/estimator output and $y=f(\mu_1)$. Hence, we can write  (\ref{eqIt}) using the following relation (see Fig. \ref{fig:sector_bound}):
\begin{equation}
I(t)dt=2\alpha(f(\mu_1)+\beta \mu_1)dt+dW(t),
\end{equation}
where $\beta$ is the slope of the tangent on the curve at $\mu_1=0$. Also, 
\begin{equation}
\begin{split}
I(t)dt&=[2\alpha(f(\mu_1)+2\alpha\beta(\phi(t)-u(t))]dt+dW(t)\\
I(t)+2\alpha\beta u(t)&=2\alpha f(\mu_1)+2\alpha\beta\phi(t)+W(t)
\end{split}
\end{equation}
and 
\begin{equation}
\begin{split}
\tilde{y}(t)&=2\alpha f(\mu_1)+2\alpha\beta\phi(t)+W(t)\\
\frac{\tilde{y}(t)}{2 \alpha \beta}&=\phi(t)+\frac{f(\mu_1)}{\beta}+\frac{1}{2\alpha\beta}W(t)
\end{split}
\end{equation}
where $\tilde{y}=I(t)+2\alpha\beta u(t)$.
Finally we can write a linear system model with sector bounded uncertainty as follows:
\begin{equation}
\label{eqyt}
\begin{split}
\dot{\phi}&=-\lambda\phi(t)dt+\sqrt{\kappa} v(t),\\
\bar{y}(t)&=\phi(t)+\frac{f(\mu_1)}{\beta}+\frac{1}{2\alpha\beta} W(t),
\end{split}
\end{equation}
where $\bar{y}(t)=\frac{\tilde{y}(t)}{2 \alpha \beta}$ We assume that the sector is defined by the region $\gamma$ (see Fig. \ref{fig:sector_bound})
\[
\gamma=\frac{1-m}{2},
\]
where $m$ is the slope of the lowest boundary of the sector. Also, $f^2(\mu_1)\leq\gamma^2 \mu_1^2$.

Let us assume that $\tilde{f}(\mu_1)={f(\mu_1)}{\gamma}$. Then (\ref{eqyt}) can be written as
\begin{equation}
\label{eqytf}
\begin{split}
\dot{\phi}&=-\lambda\phi(t)dt+\sqrt{\kappa}dv(t),\\
\bar{y}(t)&=\phi(t)+\gamma\frac{\tilde{f}(\mu_1)}{\beta}+\frac{1}{2\alpha\beta} W(t).
\end{split}
\end{equation}

\subsection{Smoother design}\label{sec:RNS}
The uncertainty model of the original system under consideration in (\ref{eqytf}) can be written in the form (\ref{eqsystem}) as follows:
\begin{equation}
\label{eqsystem1}
\begin{split}
d\phi(t)&=-\lambda\phi(t)d(t)+[0\times \tilde{f}(\mu_1)(t)\\
&+\sqrt{\kappa} \Delta_1 \zeta_1] dt +[\sqrt{\kappa} \quad 0] dW(t),\\
\nu_1(t)&=\gamma 2 \alpha \phi(t),\\
\zeta_1&=0 \times \phi(t),\\
\bar{y}(t)&=\phi(t)+2\alpha\gamma\frac{\tilde{f}(\mu_1)}{2\alpha\beta}+\frac{1}{2\alpha\beta} W(t),
\end{split}
\end{equation}
where $\Delta_1$ represents the uncertainty in system parameters. However, in this example, we only consider uncertainty in measurment Therefore, $\Delta_1=0$.  The measurement uncertainty $\tilde{f}(\mu_1)$ satisfies the IQC (\ref{eqIQC1}). A comparison of the above model with (\ref{eqsystem}) gives the following model parameters:
\[
A=-\lambda,~ \bar{B}_{11}=0, ~B_{11}=\sqrt{\kappa}, B_1=[\sqrt{\kappa} \quad 0 \quad 0],
 \]
 \[
 C_2=1, \bar{D}_{21,1}=\frac{1}{2\alpha \beta},
\]
\[
D_{21,1}=0, D_{21}=[0 \quad \frac{1}{2\alpha \beta}],~ \bar{C}_{1,1}=\gamma 2\alpha, ~ C_{1,1}=0,
\]
\[
g=1,~k=1.
\]
\begin{table}
\caption{Parameters values for the optical system.}
\label{tab2}
\centering
\begin{tabular}{|c|c||c|c|}
\hline $\lambda$ & $9.14 \times 10^{3}$ rad/s & $\gamma$ & $0.4$ \\
\hline $\kappa$ & $40000$ rad/s& $\alpha$ & $1162$ s$^{-1}$ \\
\hline $\beta$ & $1$ &  & \\
\hline
\end{tabular}
\end{table}
The values of the system parameters are given in Table \ref{tab2}. The augmented system model parameters in (\ref{eqNsystem}) using second order Pade approximation of the delay $\delta=3.1 \mu~sec$ can be obtained as follows:
\begin{small}
\begin{equation}
\begin{split}
&A_p=\left[
\begin{array}{ccc}
-9.14\times 10^{3} & 0 & 0\\
2048 & -1.94\times 10^{6} & -1.19\times 10^{6}\\
0 & 1.048\times 10^{6} & 0\\
\end{array}
\right], \\ 
&B_{p1}=\left[
\begin{array}{cc}
200&0\\
0&0\\
0&0\\
\end{array}
\right],~B_{p1i}=\left[
\begin{array}{c}
200\\
0\\
0\\
\end{array}
\right],~\bar{B}_{p1i}=\left[
\begin{array}{c}
0\\
0\\
0
\end{array}
\right],\\
&C_{p0}=\left[
\begin{array}{c}
1\\
0\\
0
\end{array}
\right]^T,~
C_{p1,1}=\left[
\begin{array}{c}
0\\
0\\
0
\end{array}
\right]^T,~
\bar{C}_{p1,1}=\left[
\begin{array}{c}
929.6\\
0\\
0
\end{array}
\right]^T,\\
&C_{p2}=\left[
\begin{array}{ccc}
1 & 0 & 0
\end{array}
\right].
\end{split}
\end{equation}
\end{small}
Considering the above model parameters, the matrices defined in (\ref{eqCsystem}) are given as follows:
\begin{small}
\begin{equation}
\begin{split}
&\tilde{B}_1=\left[
\begin{array}{ccc}
200 & 0 & 0\\
0 & 0 & 0\\
0& 0 & 0\\
\end{array}
\right]
,~ \tilde{C}_1=\left[
\begin{array}{ccc}
0 & 0 & 0\\
929.6 & 0 & 0\\
0& 0 & 0\\
\end{array}
\right],\\
&\tilde{D}_{12}=\left[
\begin{array}{cc}
0&0\\
0&0\\
0&1\\
\end{array}
\right],~
\tilde{C}_2=\left[
\begin{array}{ccc}
1 & 0 & 0\\
0 & 0 & 0\\
\end{array} \right],~\tilde{D}_{21}
=\bar{D}_{21}\\
&\bar{D}_{21}
=\left[
\begin{array}{ccc}
0 & 4\times 10^{-4}& 0\\
0 & 0 & 1\\
\end{array}
\right],~
h=1, ~ r=1,~ p=3,~ \tilde{k}=4.
\end{split}
\end{equation}
\end{small}
For the above definition of the model, assumptions (1) and (2) are satisfied for $J(t)=I$ and $0< d_0\leq 1$.
From the definition of the IQC (\ref{eqIQC}) we have
\[
M_{\lambda}(\lambda)=\left(
\begin{array}{ccc}
\lambda_1 & 0 & 0 \\
 0 & \lambda_2+\lambda_3 & -\lambda_2 \\
 0 & -\lambda_2 & \lambda_2+\lambda_4 \\
\end{array}
\right)\]
and its inverse is given by
\[
M^{-1}=\left(
\begin{array}{ccc}
 \frac{1}{\lambda_1} & 0 & 0 \\
 0 & \frac{\lambda_2+\lambda_4}{\lambda_3 \lambda_4+\lambda_2 (\lambda_3+\lambda_4)} &
   \frac{\lambda_2}{\lambda_3 \lambda_4+\lambda_2 (\lambda_3+\lambda_4)} \\
 0 & \frac{\lambda_2}{\lambda_3 \lambda_4+\lambda_2 (\lambda_3+\lambda_4)} &
   \frac{\lambda_2+\lambda_3}{\lambda_3 \lambda_4+\lambda_2 (\lambda_3+\lambda_4)} \\
\end{array}
\right).
\]
The constraints on $\lambda_i$ for $i=1,2,3,4$ due to (\ref{eqcons}) are given below:
\[
\lambda_1>0,\quad \lambda_2>0, \quad \lambda_3>0, \quad \lambda_4>0,\quad 0<\lambda_1\leq 1,
\]
\[
0< \lambda_2 +\lambda_3 \leq 1, \quad 0< \lambda_2 +\lambda_4 \leq 1,
\]
\[
\quad (1-\lambda_2-\lambda_3)(1-\lambda_2-\lambda_4)-\lambda_2^2\geq 0.
\]
For the values of the parameters given in Table \ref{tab2}, the minimum value of the bound (\ref{eqbound}) obtained using an `Interior-point' numerical optimization method is $V_{\tau}=0.15$ at $\bar{\tau}=1.13\times 10^{-6}$ and the values of the $\lambda_i$ for $i=1,2,3,4$ are obtained as follows: 
\[
\lambda_1=0.9727,~ \lambda_2=0.4831,~ \lambda_3=0.0015,~
\lambda_4=0.0014.
\]
The estimator parameters in (\ref{eqCparams}) are calculated as follows:
\begin{small}
\begin{equation}
\begin{split}
&A_c=\left[
\begin{array}{ccc}
-4.58\times 10^{5} & -0.09 & -7.14\\
1.93\times 10^{3} & -1.93\times 10^{6} & -1.19\times 10^{6}\\
-550.1& 1.0486\times 10^{6}& -0.01
\end{array}
\right], \\ 
&\tilde{B}_c=\left[
\begin{array}{cc}
4.45\times 10^{5} & -190.97\\
120.98 &-0.052\\
545.41 & -0.233\\
\end{array}
\right],\\
&\tilde{C}_c=\left[
\begin{array}{ccc}
1.02 & 4.21\times 10^{-7}& 3.23\times 10^{-5}\\
944.68 &3.9\times 10^{-4} & 0.0299\\
\end{array}
\right].
\end{split}
\end{equation}
\end{small}

\begin{figure}[t]
\hfill
\begin{center}
\epsfig{file=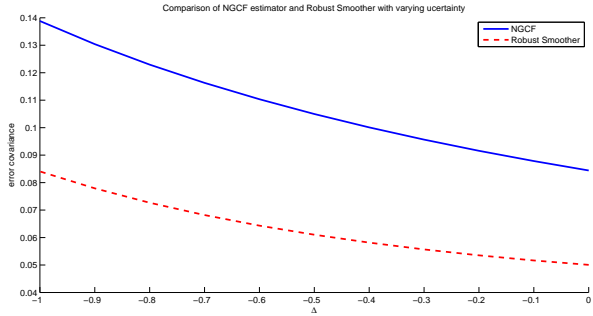, scale=0.33}
\caption{Error covariance with uncertainty variation.}
\label{fig:error_NGCF}
\end{center}
\end{figure}

\subsection{Closed loop simulation}\label{sec:sim}
In this paper, simulations with both linear and nonlinear uncertainties have been performed for $\Delta$ varying from $0$ to $-1$. The linear simulation gives an insight into the behavior of the error covariance of the system with varying uncertainty. Linear simulation also enables comparison of the nonlinear guaranteed cost filter (NGCF) with the robust smoother of this paper for different values of  uncertainty. Since the nature of the uncertainty in the given system is actually nonlinear therefore, a Monte-Carlo simulation using Simulink has also been performed for the closed loop system and result is presented in the section.
\subsection{Linear Simulation}
A linear simulation with uncertainty has been performed for the closed loop system  with and without the smoother. For this example, the matrix $\Delta$ is defined as follows:
\begin{equation}
\Delta=\left[
\begin{array}{ccc}
\Delta_1 & 0 & 0\\
0 &\Delta_2 & 0\\
0 & 0 &\Delta_2\\
\end{array}
\right],
\end{equation}
where $\Delta_2$ represents the uncertainty in measurements.
Error covariance for the system with smoother is calculated using (\ref{eqCo}). Solid blue line in Fig. \ref{fig:error_NGCF} represents the error covariance of the nonlinear guaranteed cost filter (NGCF) and red broken line represents the error covariance of robust smoother. Also, $\Delta_2=0$ means zero uncertainty and $\Delta_2=-1$ represents maximum uncertainty for which the estimator has been designed; i.e. $\gamma=0.4$. It is obvious that as the uncertainty increases, the error covariance increases for both filter and smoother. However the error covariance of smoother is much less than the NGCF and also less sensitive to the uncertainty.
\begin{remark}
The uncertainty region lies below the nominal value of $\beta=1$ (see Fig. \ref{fig:sector_bound}). Therefore $\Delta_2$ can only vary between $0$ and $-1$.
\end{remark}
\subsection{Nonlinear simulation}
Since the original model of the system is nonlinear, a nonlinear simulation has also been performed. A nonlinear Monte-Carlo simulation has been performed by collecting error samples during $50,000$ runs using Simulink. The nonlinear uncertainty is of the form $\sin(\phi)-\phi$. The samples have been collected by running the simulation for a fixed time interval of $1$ msec with sample time of $0.01~\mu$ sec and with randomly generated noise signals. The error signal is collected at the end of the simulation.  The error covariance for the simulation with the robust smoother is found to be $0.0605$. In the case of nonlinear uncertainty a similar simulation with the NGCF in the feedback loop yields an error covariance of $0.1031$ which is $41\%$ larger than the error covariance obtained using the robust smoother.
\

\section{Conclusion} \label{sec:concl}
In this paper the problem of nonlinear robust fixed lag smoothing has been considered for an uncertain system. The scheme proposed in this paper extends the design of a nonlinear guaranteed cost filter by using a continuous time fixed lag smoother. The scheme uses an uncertainty model for the system and provides a nonlinear robust smoother by considering a copy of nonlinear uncertainty in the estimator/smoother. The scheme is applied to a phase estimation problem of a quantum optical system. Simulation results with both linear and nonlinear simulations show that the robust smoother scheme improves the error covariance significantly and works better than the corresponding robust filter.  Further research will be directed toward implementing the scheme on real hardware and performing experiments.
\section{ACKNOWLEDGMENTS}
This research was supported by the Australian Research Council.


\bibliographystyle{IEEEtran}        
\bibliography{Literature_jabref}   

\begin{thebibliography}{10}
\providecommand{\url}[1]{#1}
\csname url@samestyle\endcsname
\providecommand{\newblock}{\relax}
\providecommand{\bibinfo}[2]{#2}
\providecommand{\BIBentrySTDinterwordspacing}{\spaceskip=0pt\relax}
\providecommand{\BIBentryALTinterwordstretchfactor}{4}
\providecommand{\BIBentryALTinterwordspacing}{\spaceskip=\fontdimen2\font plus
\BIBentryALTinterwordstretchfactor\fontdimen3\font minus
  \fontdimen4\font\relax}
\providecommand{\BIBforeignlanguage}[2]{{%
\expandafter\ifx\csname l@#1\endcsname\relax
\typeout{** WARNING: IEEEtran.bst: No hyphenation pattern has been}%
\typeout{** loaded for the language `#1'. Using the pattern for}%
\typeout{** the default language instead.}%
\else
\language=\csname l@#1\endcsname
\fi
#2}}
\providecommand{\BIBdecl}{\relax}
\BIBdecl

\bibitem{Frank_estimation}
F.~L. Lewis, L.~Xie, and D.~Popa, \emph{Optimal and Robust Estimation, with an
  introduction to stochastic control theory}, F.~L. Lewis, Ed.\hskip 1em plus
  0.5em minus 0.4em\relax CRC Press, 2008.

\bibitem{Meditch_smoothing}
J.~S. Meditch, ``A survey of data smoothing for linear and nonlinear dynamic
  systems,'' \emph{Automatica}, vol.~9, pp. 151--162, 1973.

\bibitem{Kailath_Survey}
T.~Kailath, ``A view of three decades of linear filtering thoery,'' \emph{IEEE
  Transaction On Information Theory}, vol.~20, no.~2, pp. 146--181, March 1974.

\bibitem{Anderson_smoother}
S.~Chirarattananon and B.~Anderson, ``Stable fixed-lag smoothing of continuous
  time process,'' \emph{IEEE Transaction On Information Theory}, vol. IT-20,
  no.~1, pp. 25--36, January 1974.

\bibitem{NGC_Petersen}
I.~R. Petersen, ``Robust guaranteed cost state estimation for nonlinear
  stochastic uncertain systems via an {IQC} approach,'' \emph{System \& Control
  Letters}, vol.~58, no.~11, pp. 865 --870, 2009.

\bibitem{IP}
I.~R. Petersen, V.~A. Ugrinovskii, and A.~V. Savkin, \emph{Robust control
  design using $\mathcal{H}^{\infty}$ methods}.\hskip 1em plus 0.5em minus
  0.4em\relax London: Springer, 2000.

\bibitem{Rehman_NGC}
O.~Rehman and I.~R. Petersen, ``Robust nonlinear estimation of varying optical
  phase,'' in \emph{Submitted to European control conference}, 2013.

\bibitem{Wheatley_Adaptive}
T.~A. Wheatley, D.~W. Berry, H.~Yonezawa, D.~Nakane, H.~Arao, D.~T. Pope, T.~C.
  Ralph, H.~M. Wiseman, A.~Furusawa, and E.~H. Huntington, ``Adaptive optical
  phaser estimation using time-symmetric quantum smoothing,'' \emph{Physical
  Review Letters}, vol. 104, no. 093601, 2010.

\bibitem{Shibdas_RF}
S.~Roy, I.~R. Petersen, and E.~H. Huntington, ``Robust filtering for adaptive
  homodyne estimation of continuously varying optical phase,'' in
  \emph{Australian Control Conference}, Sydney, Australia, 16-17 Nov 2012.

\bibitem{Rehman_RNE}
O.~Rehman, I.~R. Petersen, H.~Song, and E.~H. Huntington, ``Robust nonlinear
  estimation of varying optical phase,'' in \emph{Submitted to European Control
  Conference}, 2013.

\end{thebibliography}

\end{document}